\documentclass[aps,prd,preprint]{revtex4}
\usepackage{graphicx}
\begin{document}
\draft
\def\ds{\displaystyle}
\title{Trapped Particles in $PT$ Symmetric Theories}
\author{C. Yuce$^\star$}
\address{Department of Physics, Anadolu University,
 Eskisehir, Turkey}
 \email{cyuce@anadolu.edu.tr}
\author{A. Kurt}
\author{A. Kucukaslan}
\address{Department of Physics, Canakkale University, Canakkale,
Turkey}
\date{\today}
\begin{abstract}
$PT$ symmetric quantum mechanics for a particle trapped by the
generalized non-Hermitian harmonic oscillator potential is
studied. It is shown that energy and the expectation value of the
position operator $x$ can not be real simultaneously, if the
particle is trapped. Non-vanishing boundary conditions for the
trapped particle in $PT$ symmetric theory are also discussed.
\end{abstract}
\maketitle

\section{Introduction}

It is commonly believed that the Hamiltonian must be hermitian in
order to ensure that the energy spectrum is real and that the time
evolution of the theory is unitary. Although this axiom is
necessary to guarantee these desired properties, the recent
studies have shown that it is not sufficient
\cite{R1,R2,R3,R4,R5,R6,R7}. It is shown that there is a simpler
and more physical alternative axiom to Hermicity. It is the
space-time reflection symmetry ($PT$ symmetry). The Hamiltonans
may not be symmetric under $P$ or $T $ separately, It should be
invariant under their combined operation.\\
The first examples for complex potentials with real spectra were
found by using the numerical techniques. After the first examples,
further ones  have been identified using perturbative technique
\cite{PERT} and some exactly solvable $PT$ symmetric Hamiltonians
\cite{pot1,pot2,pot3,pot4,pot5,pot6,pot7,pot8,osc,coulomb,asil}
have been found. Exactly solvable examples are extremely useful in
the understanding of unusual features of $PT$ symmetric problems
and the underlying new physical concepts. It is a promising
development, with possible applications ranging from the field
theories \cite{FIELD1,FIELD2} to the supersymmetric models
\cite{SupSym1,SupSym2}.\\
Since $PT$ symmetry is an alternative condition to Hermicity, it
is now possible to construct infinitely many new Hamiltonians that
would have been rejected before because they are not Hermitian.\\
Let us recall the properties of $PT$ operation. The parity
operator is linear and has the effect
\begin{equation}
p\rightarrow -p,~~~~x\rightarrow -x.
\end{equation}
The time-reversal operator is antilinear and has the effect
\begin{equation}
p\rightarrow -p,~~~~x\rightarrow x,~~~~i\rightarrow -i.
\end{equation}
Note that the Heisenberg algebra $[x,p]=i$ is preserved since $T$
changes the sign of $i$. \\
In recent times, it has been found that in contrast with the first
conjectures, neither Hermicity nor $PT$ symmetry serves as a
sufficient condition for a quantum Hamiltonian to preserve the
reality of energy eigenvalues
\cite{psodo1,psodo2,psodo3,psodo4,psodo5,psodo6}. In fact, it has
been realized that the existence of real eigenvalues can be
associated with a non-Hermitian Hamiltonian provided it is
$\eta$-pseudo Hermitian $H^\dagger=\eta H \eta^{-1}$. In this
context, $PT$ symmetry is $P$-pseudo-Hermicity for one dimensional
Hamiltonians.\\
There are more than one method to study the non-Hermitian
Hamiltonian with the real spectra. One of the methods is to
generate the non-Hermitian Hamiltonian from the corresponding
Hermitian Hamiltonian by a complex shift of coordinate $\ds{x
\rightarrow x-i c}$, where $c$ is a constant. For example, with
this imaginary coordinate shift, the harmonic oscillator
Hamiltonian becomes non-hermitian Hamiltonian giving the real
energy spectrum \cite{asil}.

In this study, we aim at whether confinement of a particle leads
to the new physical implications in $PT$ symmetric theories. To
study this phenomenon, so-called complex shift of coordinate
method will be utilized. Here, our motivation is to investigate
whether the energy eigenvalues remain real if the particle is
trapped into a box, namely, the wave function which vanishes at a
finite point. The generalized harmonic oscillator potential is
chosen for our study since the
exact solutions are also well-known for this boundary conditions.\\
This paper is organized as follows. In the following section, the
exact solution for the trapped particle under the generalized
harmonic oscillator potential will be reviewed. In the last
section, the imaginary coordinate shift will be applied both to
the Hamiltonian to generate the non-Hermitian one and to the wave
function to check the reality of energy eigenvalues.

\section{Formalism}

In this section, we review the exact solution for the trapped
particle under the generalized harmonic oscillator potential
\cite{cem}. The confinement of a particle plays an important role
in the application of the quantum theory.  For example, the
electrons in a semiconductor are confined by a potential well in
1D (quantum well), 2D (quantum wire), or 3D (quantum dot). As a
first approximation, these materials can be understood from a
particle in a box perspective. Trapping it in a small space means
that only particular types of wave-function are allowed. These
wave-functions have each a particular characteristic length as
well as a particular energy. Note that as the box size reduces the
energy spacing changes. Hence one may expect that the extent of
confinement changes the spectroscopic properties of quantum dots.
Because of the wide application of confinement, $PT$ symmetric
version of confinement phenomena is worthy studying.\\
The generalized harmonic oscillator potential is of great
significance in understanding the new physical effects of $PT$
symmetric quantum theory since it can be solvable analytically. It
is of harmonic plus inverse harmonic type. The corresponding
Schrodinger equation for this potential reads
\begin{equation}\label{ilkdenk}
-\frac{\partial^2 \Psi}{\partial x^2}+\left( \omega^2 (t)
x^2+\frac{g}{x^2} \right) \Psi=i\frac{\partial \Psi}{\partial t}~,
\end{equation}
where angular frequency $ \omega^2 (t)$ is in general taken to be
time-dependent and $g$ is the coupling constant. The constants in
Schrodinger equation are set to unity for simplicity $\ds{(\frac{\hbar^2}{2m}=\hbar=1)}$.\\
Firstly, to find the exact confined solution of the problem, the
following transformation on the wave function is introduced.
\begin{eqnarray}
\Psi (x,t)= \exp{\left(~i \alpha \frac{x^2}{2}-\int_0^t
dt^{\prime} \alpha (t^{\prime})~\right)} \Phi (x,t)~.
\end{eqnarray}
where $\ds{\alpha (t)}$ is to be determined later. If we apply
this transformation into the Schrodinger equation, we get
\begin{equation}\label{101}
-\frac{\partial^2 \Phi}{\partial x^2}-2i \alpha  x \frac{\partial
\Phi}{\partial x}+\frac{g}{x^2} \Phi=i\frac{\partial
\Phi}{\partial t}~,
\end{equation}
Here, $\ds{\alpha (t)}$ is assumed to satisfy the following
relation to get rid of the quadratic term in the above equation
\begin{equation}\label{301}
\omega^2+\frac{\dot{\alpha}}{2}+\alpha^2=0~.
\end{equation}
The solution to this equation determines $\ds{ \alpha(t)}$. Now,
we aim at solving the equation (\ref{101}). To do this, the
coordinate is scaled by a time-dependent function $\ds{L(t)}$. The
relation between the coordinate $x$ and the new coordinate $q$ is
introduced as follows
\begin{equation}\label{102}
x \rightarrow q=\frac{x}{L(t)}~,
\end{equation}
with a consequent transformation for the time derivative operator
which is given by $\ds{\partial /
\partial t \rightarrow
\partial / \partial t-{(\dot{L}/L)}~ q~
\partial / \partial q}$, where dot denotes time derivation. We
have a freedom to choose $L(t)$. Let $\ds{L(t)}$ be chosen as
follows.
\begin{equation}\label{11885}
2 \alpha =\frac{\dot{L}}{L}~.
\end{equation}
Instead of dealing with the relation between the time-dependent
functions $L(t)$ and $\alpha (t)$, it is useful to study with the
relation between  $L(t)$ and $\omega^2 (t)$. It can be computed by
using the equations (\ref{301},\ref{11885})
\begin{equation}\label{1000}
4\omega^2 =-\frac{\ddot{L}}{L}~.
\end{equation}
The equation (5) become time-independent and includes only the
inverse harmonic oscillator potential
\begin{equation}\label{1ret0}
-\frac{\partial^2 \Phi (q)}{\partial q^2}+\left(\frac{g}{q^2}
\right) \Phi (q)=E\Phi (q).
\end{equation}
In the last step, we used the separation of variables technique.
It is given by
\begin{eqnarray}
\Phi (q,t)= \exp{(-i \int_0^t dt^{\prime} \frac{E}{L^2}~)} \Phi
(q)~,
\end{eqnarray}
where $E$ is the constant and it does not coincide with the energy
eigenvalues in general since $L(t)$ is time-dependent in general.\\
The equation (\ref{1ret0}) is a well known equation in physics and
it's solution can be found easily. It is given by
\begin{equation}\label{552}
\Phi(q)= q^{1/2} J_{\nu} (\sqrt{E}~q)~,
\end{equation}
where $\ds{\nu= \frac{1}{2} (1+4g  )^{1/2}}$ and $J_v$ is the
Bessel function. The confined solution is given in term of the
Bessel function which vanishes at some points. Transforming
backwards yields the exact wave function for the generalized
harmonic oscillator
\begin{equation}\label{2315}
\Psi_{\nu}(x,t)=N \exp{\left(~i \alpha \frac{x^2}{2}-i\int_0^t
dt^{\prime}   \frac{E}{L^2}~\right)} \frac{1}{\sqrt{L}}~R_{\nu}
(x,t) ~,
\end{equation}
where $N$ is a normalization constant and $\ds{ R_{\nu} (x,t)}$ is
given by
\begin{equation}\label{956}
R_{\nu}(x,t)=(\frac{x}{L})^{1/2} J_{\nu} (\sqrt{E} ~\frac{x}{L})~.
\end{equation}
In the equation (\ref{2315}) we used the relation,
$\ds{\frac{1}{\sqrt{L}}=\exp{(-\int_0^t \alpha dt^{\prime} )}}$ by
(\ref{11885}). Bessel function is used in the solution of the
generalized harmonic oscillator potential. Since Bessel function
is zero at some finite points, it describes the trapped particle
from the physical point of view. Now, let us make some remarks on
the boundary conditions. Since Bessel function includes the time
dependent function $L(t)$, the boundary condition is not
stationary anymore. In general, it is a moving boundary condition.
Bessel function in the equation (\ref{956}) represents trapped
particle into the box whose wall moves. For example, if
$\ds{\omega^2 (t)}$ is constant, then $L(t)$ is sinusoidal by
(\ref{1000}).\\
Having obtained the exact solution, let us write the new boundary
conditions.
\begin{equation}\label{103}
\Psi(x=L(t),t)=0 \Rightarrow \Phi(q=1,t)=0~;~~~~~~\Psi(x=0,t)=0
\Rightarrow \Phi(q=0,t)=0~.
\end{equation}
The time-dependent function $\ds{L(t)}$ which determines boundary
of the trap can not be chosen completely free. It depends on the
time dependent character of $\ds{\omega^2 (t)}$. As can be seen
from the boundary condition (\ref{103}), the particle is confined
into a box whose wall moves with the speed $\ds{\dot{L} (t)}$. To
satisfy the boundary condition (\ref{103}), the constant $E$ takes
discrete values. These values are the roots of the Bessel function
$\ds{J_{\nu} (\sqrt{E} )} $ for a given value of $\nu$.\\
Although the exact solution can be constructed only for some
special choice of $L(t)$, it's implications turns out to be quite
useful in understanding of $PT$ symmetric quantum theory as can be
seen in the next section.\\
Having obtained the exact solutions corresponding the two
different boundary conditions, let us apply the complex shift of
coordinate method to study $PT$ symmetric extension of generalized
harmonic oscillator problem.

\section{Complex Shift of Coordinate}

It is of great importance to understand the underlying new
physical concepts of $PT$ symmetric quantum mechanics. In the
past, any non-hermitian Hamiltonian was omitted, because it was
thought that the corresponding energy eigenvalues are not real. In
recent years, the reality of energy spectrum for some
non-Hermitian Hamiltonians have been shown. The reality of energy
eigenvalues of the trapped particle for a non-Hermitian
Hamiltonian has not been investigated up to now.\\
It may be worthwhile to look for some analytically solvable $PT$
symmetric potentials for the study of reality of energy spectrum
of the trapped particle with a non-hermitian Hamiltonian. One of
such potentials can be generated by a complex shift of coordinate.\\
\begin{equation}\label{201}
V(x)=\omega^2(t)~ (x-ic)^2+\frac{g}{(x-ic)^2}~.
\end{equation}
Shifting the coordinate from $x$ to $z=x-i c$, removes the
singularities on the real line, and extends the potential from the
half line to the full line. Now, our aim is to investigate the
reality of the energy spectrum for this non-Hermitian Hamiltonian.
Znojil proved that the energy eigenvalues are real
\cite{asil,modified} provided that the exact wave function of
(\ref{201}) which vanishes at infinity is used. The eigenvalues
and the eigenfunctions of the generalized harmonic oscillator with
$\ds{\omega^2=1}$ is given by \cite{asil}
\begin{eqnarray}\label{202}
\Psi_{nq}&=&N z^{-q\beta+1/2}~\exp{(-\frac{z^2}{2})}  L_n^{-q\beta} (z^2)~; \\
E_{nq}&=&4n+2-2q\alpha ~,
\end{eqnarray}
where $L_n^{\beta}$ are the associated Laguerre polynomials,
$\ds{(\beta^2-1/4=g)}$,
$n=0,1,2,...$ and $q=\pm 1$ is the quasi parity.\\
The complex extension of the harmonic plus the inverse harmonic
potential is of great importance, since it's energy eigenvalues
are real as can be seen in the equation (19). However, it is not
clear whether the energy eigenvalues remain real for the particle
trapped in a box for the non-Hermitian potential
(\ref{201}).\\
The confined solution of (\ref{201}) to the equation is obtained
from the equation (\ref{2315}) by shifting the coordinate
\begin{equation}\label{eq20}
\Psi_{\nu}(x-ic,t)= N \exp{\left(~i \alpha \frac{(x-i
c)^2}{2}-i\int_0^t dt^{\prime} \frac{E}{L^2}~\right)}
\frac{1}{\sqrt{L}} R_{\nu} (x-ic,t) ~.
\end{equation}
This solution can be compared to the another solution (\ref{202})
which vanishes at infinity.\\
For a better understanding, let us use a special value of $\nu$ in
the Bessel function. If we set $\ds{\nu=1/2}$
\begin{equation}\label{3b12}
\sqrt{\frac{x-ic}{L}}~ J_{1/2} (\sqrt{E}~\frac{x-ic}{L})=
\sqrt{\frac{2}{E\pi}} \sin (\sqrt{E}~\frac{x-ic}{L})~.
\end{equation}
Before investigating the reality of energy spectrum for this wave
function, let us check the boundary condition if we shift the
coordinate.\\
The wave function of the physically trapped particle goes to zero
at the imaginary points in this case.
\begin{equation}
\Psi_{\nu}(ic,t)=\Psi_{\nu}(L+ic,t)=0~,
\end{equation}
Shifting the coordinate from $x$ to $z=x-i c$ changes also the
boundary condition. The new model with complex boundary conditions
to describe the trapped particle in non-hermitian quantum
mechanics is different from the model used to describe the trapped
particle in standard quantum mechanics. However, the reality of
the length of the box (quantum dot) is not violated under complex
shift of coordinate. It is imposed by the nature that the length
of the box must be real like the energy spectrum. In other words,
not only the energy eigenvalues but also the length of, say,
quantum dot must be real. Although the wave function is set to
zero at the complex points describing the locations of the wall of
the box, the length of the box is still physical, namely,
$\ds{\Delta L=L-0\rightarrow(L+ic)-(0+ic)=L}$. The reality of the
length of
the box is preserved with the less strict boundary conditions.\\
To get a deep understanding of physical implications of
confinement in $PT$ symmetric theory, let us find the absolute
square of the wave function by using (\ref{3b12})
\begin{eqnarray}\label{156l}
|R_{1/2}|^2&=& \frac{2}{E \pi} ~\sin (\sqrt{E} ~\frac{x-i c}{L})
\sin (\sqrt{E}
~\frac{x+i c}{L})\nonumber \\
&=&\frac{1}{E \pi} ~\left( ~\cosh(~\frac{2\sqrt{E}}{L}
~)-\cos(~\frac{2\sqrt{E}}{L} x ~) \right) ~.
\end{eqnarray}
Trapping of a particle can be modeled using particle in a box
perspective. From the experimental point of view, not the wave
function $\Psi$ but it's absolute square $|\Psi|^2$ is important.
Actually, the probability of finding the particle can be measured
at only real points. If we look at (\ref{156l}), we observe that
the absolute square of the wave function is not zero at the walls
of the box $\ds{(x=0, x=L)}$. It leads to the non-vanishing
boundary conditions for $|\Psi|^2$.
\begin{equation}
|\Psi|^2_{x=0,  ~L} \neq 0 ~.
\end{equation}
This result has crucial implications. A consistent picture of
confinement in $PT$ symmetric quantum theory leads to unexpected
result. In usual quantum theory, both the wave function $\Psi$ and
it's absolute square $|\Psi|^2$ vanish at the wall of the box.
When making measurements to find the probability of a trapped
particle at the wall of the box, the two theory gives different
result.\\
So far, the case $\ds{\nu=1/2}$ have been studied for a better
understanding the physical implications. The validity of these
results can also be checked for different values of $\ds{\nu}$ .
\begin{eqnarray}
|R_{\nu}|^2=\sqrt{\frac{x^2+c^2}{L}} ~J_{\nu} (\sqrt{E}~
\frac{x-ic}{L})~ J_{\nu} (\sqrt{E} ~\frac{x+ic}{L})~.
\end{eqnarray}
Having investigated the boundary conditions, let us now find a
relation for the expectation values for the trapped
particle in the non-hermitian harmonic oscillator potential. \\
Instead of calculating the energy eigenvalues directly, our
strategy is to start from the energy eigenvalues for the Hermitian
Hamiltonian (\ref{ilkdenk}) in order to investigate the reality of
energy eigenvalues for the non-Hermitian one.
\begin{equation}\label{eq2615}
\int_{0}^{L} \Psi^{\star}(x) H(x) \Psi(x) dx  \in \Re~,
\end{equation}
We know that the energy eigenvalues are real for the Hermitian
harmonic oscillator potential given by (\ref{ilkdenk}) from which
the corresponding non-Hermitian Hamiltonian is obtained by the
complex shift of the coordinate. If we make a change of variable
$\ds{x\rightarrow x-i c}$ in the above integral, the result of the
integration is not changed. Hence, the reality of the integration
is not lost
\begin{equation}
\int_{-ic}^{L-ic} \Psi^{\star}(x-ic) H(x-ic) \Psi(x-ic) dx ~.
\end{equation}
The wave function was chosen to be zero at $\ds{x=ic,L+ic}$ (21).
To reach a true expression for the expectation value, let us take
the complex conjugate of the above equation.
\begin{equation}
\int_{ic}^{L+ic} \Psi^{\star}(x-ic) H(x+ic) \Psi(x-ic) dx~,
\end{equation}
where we have used $\ds{H^{\star}(x-ic)=H(x+ic)}$. We need the
relation between $H(x+ic)$ and $H(x-ic)$ to proceed. As a special
case, let us find this relation for the case with pure
non-Hermitian harmonic oscillator potential $(g=0)$. It is given
by $\ds{ H(x+ic)=H(x-ic)+4i\omega^2 cx }$. Substituting this into
the abobe equation, we obtain
\begin{equation}\label{eq30}
\int_{ic}^{L+ic} \{ \Psi^{\star}(x-ic)  \left(H(x-ic)+4i\omega^2
cx\right)
 \Psi(x-ic)\} dx  ~.
\end{equation}
\begin{equation}\label{efgndfhgbdfs}
<E>+4i\omega^2 c~ <x> \in \Re~.
\end{equation}
It is a remarkable result that energy and the expectation value of
the position operator $\ds{x}$ can not be real simultaneously for
the trapped particle if $\omega$ is not zero. The reality of
energy eigenvalues is preserved for the trapped particle with the
non-Hermitian Hamiltonian only for free particle
$\ds{\omega^2=0}$. This is reasonable since the free particle
Hamiltonian remains still Hermitian if we shift the coordinate to
the complex plane. For the non-vanishing $\omega$, the energy
eigenvalues are not real. This is an unexpected result because
they are real when the particle is not trapped for the same
non-hermitian Hamiltonian \cite{asil}.

To sum up, the generalized harmonic oscillator plays an important
role in understanding the new physical effects of $PT$ symmetric
quantum mechanics. We observed the non-vanishing probability
density at the wall of the box surrounding the trapped particle.
Furthermore, it was shown that energy and the expectation value of
$x$ can not be real simultaneously.

\end{document}